\documentclass[a4paper,aps,twocolumn]{revtex4}
\RequirePackage[english]{babel}
\RequirePackage[latin1]{inputenc}
\RequirePackage[T1]{fontenc}
\RequirePackage{mathrsfs}
\RequirePackage{amsmath}
\RequirePackage{amssymb}
\RequirePackage{amsbsy}
\RequirePackage{bm}
\begin{document}
\title{\bf{A classical calculation of the leptonic magnetic moment}}
\author{Luca Fabbri}
\affiliation{DIME, Universit\`{a} di Genova, Piazzale Kennedy Pad.D, 16129 Genova, ITALY}
\date{\today}
\begin{abstract}
In this paper we will show that purely classical concepts based on a few heuristic considerations about extended field configurations are enough to compute the leptonic magnetic moment with corrections in $\alpha$-power perturbative expansion.
\end{abstract}
\maketitle
\section{HISTORICAL INTRODUCTION}
1947's Shelter Island's conference is known to have raised awareness around two effects, the anomalous magnetic moment of the electron \cite{Schwinger:1948iu} and the Lamb shift in hydrogen atoms \cite{Lamb:1947zz,Bethe:1947id}, which were shortly later explained in terms of $n$-loop contributions; the following year another effect, the Casimir effect \cite{Casmir:1947hx}, was explained in terms of zero-point energies: as the former two can be interpreted by means of creation and re-absorption of virtual particles, while the latter one can be interpreted by means of vacuum fluctuations, these three effects are altogether taken as proof to justify the interpretation for which quantum fluctuations are real \cite{p-s} --- nevertheless, if we wish to avoid meaningless interpretations, it is necessary to take into account these figures rather gingerly.

A first instance indicating that quantum fluctuations, and more precisely vacuum fluctuations, are not necessarily real should have been seen already in the seminal paper about the Casimir effect: Casimir calculated the pull of two plates using zero-point energies only after Bohr suggested to follow a method simpler than the original one, in terms of which Casimir and Polder calculated the attraction between paired conductors employing retarded van der Waals forces; the Casimir force has also been computed in terms of radiative processes connected to external legs by Jaffe \cite{j}, or fields in interaction with external sources by Schwinger \cite{s}. In none of these calculations is the zero-point energy found: if Jaffe replaced the ground-state with higher-order corrections but always dealing with operators while Schwinger replaced operators with sources in a formalism that was essentially a path-integral formulation, Casimir and Polder considered no quantum concept. So we may ask if also for the other two effects field quantization may be avoided.

The reality of quantum fields could also be questioned in view of what might give rise to the Lamb shift for the hydrogen atom, with the hyper-fine splitting described by Bethe in terms of the quantum setting: nevertheless, it has also been shown in a semi-classical treatment by Welton that it is possible to describe these splittings as differences of the potential due to oscillations in the position of the electron \cite{Welton:1948zz}; moreover, we can make entirely classical this description if the displacement in the location of the electron is due its Zitterbewegung \cite{g}.

That the hyper-fine splittings could be re-interpreted by assuming that electrons have a trembling motion is important for the fact that in their decay rates, para-positronium and ortho-positronium display a discrepancy in the fine-structure constant \cite{p-s}: the reason may be that in the case of positronium, electrons and positrons have an elementary dynamics, but singlet and triplet states of positronium may receive different contributions if the electron and positron were to have non-trivial dynamics.

Finally, also the electron magnetic moment correction has been calculated in terms of Zitterbewegung \cite{k}.

This is important because the calculated and measured values of the anomalous magnetic moment, if in the case of the electron they agree, in the case of the muon they disagree for $3.4$ standard deviations \cite{Hagiwara:2006jt}: this discrepancy might be quenched for leptons of finite extension.

That such corrections have something to do with a finite extension is clear since the most precise tests of QED strongly depend on the precision about the measurement of the Compton wave-length of particles \cite{p-s}.

As it is clear, precision tests of QED do show discrepancies between experiments and theory.

Still worse is the fact that QED is known to have problems in its theoretical structure: the most well-known and important is that (for the energy shifts and the anomalous magnetic moment of leptons) calculations are done by using a cut-off that is not intrinsic to QED, therefore suggesting that the theory would have to fail beyond a certain energy scale; also (in the case of the anomalies for the magnetic moment of leptons) calculations are based on perturbative expansions which, despite being finite term-by-term, do not converge in the entire series.

However, an additional requirement in terms of which all calculations are done is the existence of expressions
\begin{eqnarray}
&A_{I}\!=\!UA_{0}U^{-1}\label{intoperator}\\
&|I\rangle\!=\!U|0\rangle\label{intstate}
\end{eqnarray}
which spell that operators and states in interaction are unitarily equivalent to the corresponding operators and states in free case: expressions (\ref{intoperator}-\ref{intstate}) are known as interaction picture, but they cannot hold in a Lorentz-covariant quantum field theory as proven by Haag theorem \cite{Haag:1955ev}.

As a consequence, Haag theorem, demonstrating that the interaction picture is inconsistent, tells that quantum field theory may make no sense whatsoever \cite{s-w}.

This suggests that it is preferable to employ the formalism of field theory assuming no additional arbitrary tool, and in any case it is merely a question of general interest studying what is the extent of classical fields.

In clarifying the reach of classical fields one may get an insight about the essence of quantum effects and indeed there may be a physical meaning that can be extracted from the quantization protocols: the fact that quantization prescriptions, being related to stochastic processes, involve an infinite number of degrees of freedom, makes it possible to interpret the particle as an extended field.

Assuming classical fields with a finite size and calculating what are the consequences for the leptonic magnetic moment correction is what we will do in this paper.
\section{CLASSICAL EXTENDED FIELDS}
In the introduction, we have re-called and high-lighted that quantum electrodynamics in its fundamental structure may be not well defined at all \cite{s-w}, and consequently it is wise to do calculations avoiding any form of field quantization prescription; also we have remarked that the presence of the Compton wave-length of the particle is ubiquitous \cite{p-s}, which suggests that considering point-like particles is restrictive: a leptonic magnetic moment correction up to the lowest-order was calculated for a classical particle looking like an extended field because of its Zitterbewegung in \cite{k}, although in this paper there are again additional arbitrary assumptions maybe avoidable for a classical particle as an extended field distribution.

In this direction, there are works in settings that are semi-classical \cite{Bennett:2013qla}, or classical \cite{Fabbri:2013isa,Fabbri:2014zya}, where the leptonic magnetic moment correction is shown to be present, although the generality of the treatment forbade the computation of its magnitude; nevertheless while in the previous works exact solutions were needed, here a few simple features of the material distribution would be enough.

To see what these properties are, we start from the fundamental observation that, despite the field is fundamental, nevertheless it is not irreducible: $\frac{1}{2}$-spin spinors have two complementary parts, the left-handed and the right-handed semi-spinor projections; if we think at these two components as wave-packets localized in two regions, then the two peaks are separated by a distance equalling the size of the particle given by the Compton wave-length that is associated to the mass of the particle itself.

The model we employ is a $(1\!+\!3)$-dimensional spacetime with torsion and metric forming the metric-compatible connection, orthonormal frames allow general coordinate transformations to become specific Lorentz transformations, vector $A_{\mu}$ is the electrodynamic potential, and by writing the Lorentz transformation in complex representation, Clifford matrices $\{\gamma^{\mu},
\gamma^{\nu}\}\!=\!2g^{\mu\nu}\mathbb{I}$ are introduced so that the $\frac{1}{2}$-spin spinors $\psi$ are defined; in terms of partial derivatives $F_{\mu\nu}\!=\!\partial_{\mu}A_{\nu}
\!-\!\partial_{\nu}A_{\mu}$ is the electrodynamic strength, whereas in terms of the most general covariant derivative of the spacetime and with the electrodynamic potential $D_{\mu}\psi
\!=\!\nabla_{\mu}\psi\!+\!iqA_{\mu}\psi$ is the gauge-covariant derivative with charge $q$ of the spinor field: this is the kinematic structure of the leptonic matter field we study.

For the dynamical evolution we may assume that torsion be negligible and the metric be flat although we retain the use of curvilinear coordinates; for the electrodynamic interaction of the spinor field the action is hence
\begin{eqnarray}
&\mathcal{L}\!=\!-\frac{1}{4}F_{\mu\nu}F^{\mu\nu}
\!+\!\frac{i}{2}\left(\overline{\psi}\gamma^{\mu}D_{\mu}\psi
\!-\!D_{\mu}\overline{\psi}\gamma^{\mu}\psi\right)\!-\!m\overline{\psi}\psi
\end{eqnarray}
in terms of the mass $m$ of the matter field, and whose variation gives the ensuing field equations according to
\begin{eqnarray}
&\nabla_{\alpha}F^{\alpha\nu}\!=\!q\overline{\psi}\gamma^{\nu}\psi\\
&i\gamma^{\mu}D_{\mu}\psi\!-\!m\psi\!=\!0
\end{eqnarray}
or in an equivalent explicit form that is given by
\begin{eqnarray}
&\nabla_{\alpha}\nabla^{\alpha}A^{\nu}\!=\!q\overline{\psi}\gamma^{\nu}\psi\label{e}\\
&i\gamma^{\mu}\nabla_{\mu}\psi\!-\!qA_{\mu}\gamma^{\mu}\psi\!-\!m\psi\!=\!0\label{s}
\end{eqnarray}
in Lorentz-gauge $\nabla A\!=\!0$ as known: the system of the Maxwell field equations (\ref{e}) has solution given by
\begin{eqnarray}
A^{\nu}\!=\!\frac{q}{4\pi}
\int\frac{\overline{\psi'}\gamma^{\nu}\psi'}{|\vec{r}\!-\!\vec{r}\,'|}d^{3}r'
\label{pot}
\end{eqnarray}
called retarded potentials and which can be substituted into the Dirac field equation (\ref{s}) so that they become
\begin{eqnarray}
i\gamma^{\mu}\nabla_{\mu}\psi\!-\!\frac{q^{2}}{4\pi}\!\!
\int\frac{\gamma^{\mu}\psi\overline{\psi'}\gamma_{\mu}\psi'}{|\vec{r}\!-\!\vec{r}\,'|}d^{3}r'\!-\!m\psi\!=\!0
\label{f}
\end{eqnarray}
where $\psi'\!=\!\psi'(t\!-\!|\vec{r}\!-\!\vec{r}\,'|,\vec{r}\,')$ and $\psi\!=\!\psi(t,\vec{r}\,)$ were used and showing the retardation acting on the spinor field.

From the Dirac field equation we get the Gordon form
\begin{eqnarray}
\nonumber
&\nabla_{\mu}(\frac{i}{4}\overline{\psi}[\gamma^{\alpha},\gamma^{\mu}]\psi)
\!+\!\frac{i}{2}(\overline{\psi}\nabla^{\alpha}\psi\!-\!\nabla^{\alpha}\overline{\psi}\psi)-\\
&-qA^{\alpha}\overline{\psi}\psi\!-\!m\overline{\psi}\gamma^{\alpha}\psi\!\equiv\!0
\label{dec}
\end{eqnarray}
decomposed into the divergence of the leptonic spin plus the current the lepton would have had if it were a scalar plus the retarded potentials plus the leptonic current.

In non-relativistic limit, potentials (\ref{pot}) have the purely spatial part that is given according to the expression
\begin{eqnarray}
\vec{A}\!=\!\frac{q}{4\pi}\!
\int\!\frac{\overline{\psi'}\vec{\gamma}\psi'}{|\vec{r}\!-\!\vec{r}\,'|}d^{3}r'
\label{a}
\end{eqnarray}
with no retardation after all and as it is known in the standard representation the spinor reduces to the form that is given according to $\overline{\psi}\!=\!(\phi^{\dagger},0)$ with a dependence of the type $\phi'\!=\!\phi'(\vec{r}\,')$ and $\phi\!=\!\phi(\vec{r}\,)$ themselves showing no sign of retardation; as a consequence, the Gordon decomposition (\ref{dec}) has the purely spatial part given by
\begin{eqnarray}
&\vec{\nabla}\!\!\times\!\!(\phi^{\dagger}\vec{\frac{\sigma}{2}}\phi)
\!-\!\frac{i}{2}(\phi^{\dagger}\vec{\nabla}\!\phi\!-\!\vec{\nabla}\!\phi^{\dagger}\phi)
\!-\!\phi^{\dagger}\phi q\vec{A}\!\equiv\!m\overline{\psi}\vec{\gamma}\psi
\label{j}
\end{eqnarray}
where the current the lepton would have if it were a scalar accounts for the linear momentum and so it is negligible.

Plugging solution (\ref{a}) into expression (\ref{j}) we get
\begin{eqnarray}
\vec{\nabla}\!\!\times\!\!(\phi^{\dagger}\vec{\frac{\sigma}{2}}\phi)\!\equiv\!
m\overline{\psi}\vec{\gamma}\psi\!+\!\frac{q^{2}}{4\pi}\frac{1}{m}
\!\int\!\frac{m\overline{\psi'}\vec{\gamma}\psi'\phi^{\dagger}\phi}{|\vec{r}\!-\!\vec{r}\,'|}
d^{3}r'
\label{form}
\end{eqnarray}
where the electrodynamic potentials have disappeared from an expression that now yields the curl of the spin density in terms of the mechanical momentum density plus the mechanical momentum density induced by the electrodynamic interaction of the entire field distribution.

Up to now all is general, but from this moment on we may take advantage of the heuristic interpretation of extended fields given above: in it the lepton is described in terms of two wave-packets separated in average by the Compton wave-length; the two wave-packets are to be identified with the left-handed and right-handed semi-spinor components localized in $\vec{r}$ and $\vec{r}\,'$ so that they can be written as in $\psi^{\dagger}\!=\!(L^{\dagger},0)$ and $\psi'^{\dagger}\!=\!(0,R^{\dagger})$ respectively, with the condition $|\vec{r}\!-\!\vec{r}\,'|
\!=\!\lambda$ and $\lambda$ being the Compton wave-length associated to the mass of the particle.

When the field equations (\ref{f}) are decomposed in terms of the left-handed and right-handed semi-spinorial components, we get the pair of coupled equations
\begin{eqnarray}
\nonumber
i\gamma^{\mu}\nabla_{\mu}L\!-\!\frac{q^{2}}{4\pi}\!\!
\int\frac{\gamma^{\mu}L\overline{L'}\gamma_{\mu}L'}{|\vec{r}\!-\!\vec{r}\,'|}d^{3}r'-\\
-\frac{q^{2}}{4\pi}\!\!
\int\frac{\gamma^{\mu}L\overline{R'}\gamma_{\mu}R'}{|\vec{r}\!-\!\vec{r}\,'|}d^{3}r'
\!-\!mR\!=\!0
\label{L}\\
\nonumber
i\gamma^{\mu}\nabla_{\mu}R
\!-\!\frac{q^{2}}{4\pi}\!\!
\int\frac{\gamma^{\mu}R\overline{L'}\gamma_{\mu}L'}{|\vec{r}\!-\!\vec{r}\,'|}d^{3}r'-\\
-\frac{q^{2}}{4\pi}\!\!
\int\frac{\gamma^{\mu}R\overline{R'}\gamma_{\mu}R'}{|\vec{r}\!-\!\vec{r}\,'|}d^{3}r'
\!-\!mL\!=\!0
\label{R}
\end{eqnarray}
in which it is possible to see that each of the two chiral projections has a self-interaction plus an interaction with the other chiral projection: for the self-interaction, the integral contains the pole $\vec{r}\!\rightarrow\!\vec{r}\,'$ but since in that point we also have 
$L\!\rightarrow\!L'$ and $R\!\rightarrow\!R'$ and because we know that $\gamma^{\mu}L\overline{L}\gamma_{\mu}L\!\equiv\!0$ and 
$\gamma^{\mu}R\overline{R}\gamma_{\mu}R\!\equiv\!0$ then the self-interaction terms vanish; for the mutual interaction, the distance is fixed at the Compton wave-length and therefore ultra-violet divergences do not necessarily occur.

In the Gordon decomposition the splitting in left-handed and right-handed semi-spinorial components has the same features; thus when in (\ref{form}) we split the integral in the two regions occupied by the two matter distributions that correspond to the two chiral projections we may neglect the integral containing the pole while in the remaining integral condition $|\vec{r}\!-\!\vec{r}\,'|
\!=\!\lambda$ may be used.

In the non-relativistic approximation, in chiral representation the left-handed and right-handed projections tend to become identical, implying the validity of the relationships $m\overline{\psi'}\vec{\gamma}\psi'\!\approx\!-m\overline{\psi}\vec{\gamma}\psi$ and $\phi'^{\dagger}\phi'\!\approx\!\phi^{\dagger}\phi$ as it is clear since the two opposite helicity states must have opposite spatial momentum densities and for wave-packets we have $m\overline{\psi'}\vec{\gamma}\psi'\!=\!\phi'^{\dagger}\phi'\vec{p}\,'$ and $m\overline{\psi}\vec{\gamma}\psi\!=\!\phi^{\dagger}\phi \vec{p}$ as it can be seen by employing plane waves: expression (\ref{form}) becomes
\begin{eqnarray}
\vec{\nabla}\!\!\times\!\!(\phi^{\dagger}\vec{\frac{\sigma}{2}}\phi)\!\equiv\!
m\overline{\psi}\vec{\gamma}\psi\!-\!\frac{q^{2}}{4\pi}\frac{1}{m\lambda}
\!\int\!m\overline{\psi}\vec{\gamma}\psi\phi'^{\dagger}\phi'd^{3}r'
\end{eqnarray}
and because it is always possible to have the wave-packet normalized to unity then we may finally write
\begin{eqnarray}
\vec{\nabla}\!\!\times\!\!(\phi^{\dagger}\vec{\frac{\sigma}{2}}\phi)\!\approx\!
m\overline{\psi}\vec{\gamma}\psi\left(1\!-\!\frac{\alpha}{2\pi}\right)
\label{formula}
\end{eqnarray}
since $\alpha\!=\!\frac{q^{2}}{4\pi}$ and as $m\lambda\!=\!2\pi$ is by definition the Compton wave-length associated to the mass of the particle and giving the curl of the spin density in terms of the momentum density in a very simple relationship indeed.

This expression will be used to calculate in terms of the spin the leptonic magnetic moment of the particle.
\section{LEPTONIC MAGNETIC MOMENT CORRECTION}
We may now proceed to the actual computation of the leptonic magnetic moment and its principal corrections.

The general definition of the magnetic moment comes directly from similarly general considerations about the multi-pole expansion in electrodynamics and it is
\begin{eqnarray}
\vec{\mu}\!=\!\frac{1}{2}\!\int\!\vec{r}\!\times\!(q\overline{\psi}\vec{\gamma}\psi)\,d^{3}r
\end{eqnarray}
in terms of the leptonic current; then (\ref{formula}) inverted as
\begin{eqnarray}
m\overline{\psi}\vec{\gamma}\psi\!=\!\left(1\!-\!\frac{\alpha}{2\pi}\right)^{-1}\vec{\nabla}\!\times\!(\phi^{\dagger}\vec{\frac{\sigma}{2}}\phi)
\end{eqnarray}
furnishes the form of the leptonic current in terms of the leptonic spin density: together they give the relationship
\begin{eqnarray}
\vec{\mu}\!=\!\frac{q}{2m}\left(1\!-\!\frac{\alpha}{2\pi}\right)^{-1}\!\int\!\vec{r}\!\times\!\left[\vec{\nabla}\!\times\!(\phi^{\dagger}\vec{\frac{\sigma}{2}}\phi)\right]\,d^{3}r
\end{eqnarray}
which have to be integrated over the occupied volume.

Because it is $\frac{1}{2}\vec{r}\!\times\![\vec{\nabla}\!\times\!
(\phi^{\dagger}\vec{\frac{\sigma}{2}}\phi)]\!\equiv\!\phi^{\dagger}\vec{\frac{\sigma}{2}}\phi$ up to surface terms that can be neglected inside the integral then
\begin{eqnarray}
\vec{\mu}\!=\!\frac{q}{2m}\left(1\!-\!\frac{\alpha}{2\pi}\right)^{-1}\!2\!\!\int\!
\phi^{\dagger}\vec{\frac{\sigma}{2}}\phi\,d^{3}r
\end{eqnarray}
in which the integral of the leptonic spin density is the leptonic spin: so we may write the final form
\begin{eqnarray}
\vec{\mu}\!=\!\frac{q}{2m}\left(1\!-\!\frac{\alpha}{2\pi}\right)^{-1}\!2\vec{s}
\end{eqnarray}
of the leptonic magnetic moment with the leptonic spin.

Because of the smallness of the constant we expand
\begin{eqnarray}
\vec{\mu}\!\approx\!\vec{s}\frac{q}{2m}2\left(1\!+\!\frac{\alpha}{2\pi}\right)
\end{eqnarray}
to the lowest-order of $\frac{\alpha}{2\pi}$ in the perturbative series.

According to this last formula, it is possible to read that the leptonic magnetic moment is given by the spin, times the factor $\frac{q}{2m}$ as it should be expected, times the gyro-magnetic factor $2(1+\frac{\alpha}{2\pi})$ itself being the product of the factor $2$ that recovers the prediction of the Dirac theory and the fine structure factor $1\!+\!\frac{\alpha}{2\pi}$ in which the unity is corrected by the factor $\frac{\alpha}{2\pi}$ in perturbation that recovers the prediction from Schwinger's calculations.

In our heuristic interpretation the gyro-magnetic factor has this meaning: factor $2$ comes from the two-fold multiplicity of $\frac{1}{2}$-spin spinors; in the fine-structure factor, the unity term is determined by the mechanical moment and it is corrected by powers of $\alpha$ as result of the mutual electrodynamic interaction between left-handed and right-handed semi-spinorial components: more precisely, the lowest-order power is given by $\frac{\alpha}{2\pi}$ as the result of the fact that each semi-spinorial component is a charged field moving in the electrodynamic potential induced by the other semi-spinorial component at the distance of the Compton wave-length; higher-order powers would be due to the fact that each semi-spinorial component moves in the electrodynamic potential induced by the other semi-spinorial component which itself moves in the electrodynamic potential induced by the initial semi-spinorial component, so that each would respond through the other to the action produced by itself in a back-reaction of electrodynamic origin which has the effect of changing the distance separating the two semi-spinorial projections.

We have not accounted for these corrections because if we were to consider them then in the electrodynamic self-coupling the retarded potentials could not be approximated as instantaneous and the non-relativistic limit would no longer be valid; as for the moment we shall not consider these relativistic corrections because they are beyond the aim of the present paper, but instead we will try to argue in what way a slight change in the average separation between the two semi-spinorial components would affect the result of the previous computations.

We are going to speculate on this in next section.
\section{SPECULATIVE REMARKS}
In view of further work, it is necessary to ask how we can extend these results, and clearly the answer is that we have to renounce to the single hypothesis that has been assumed in the paper, that is the fact that the extension of the matter field be the Compton wave-length exactly.

Nevertheless, it is an observed fact that the scale at which the particle starts to display wave properties is the Compton scale, and thus we are allowed to assume that the extension of the field be not much different from the Compton wave-length: if the extension of the field were given by an expression that is the Compton wave-length plus a small correction $|\vec{r}\!-\!\vec{r}\,'|
\!=\!\lambda\!+\!\delta\lambda$ then
\begin{eqnarray}
\vec{\mu}\!\approx\!\vec{s}\frac{q}{2m}2
\left(1\!-\!\frac{\alpha}{2\pi\!+\!m\delta\lambda}\right)^{-1}
\end{eqnarray}
which we can expand in powers of both $\alpha$ and $\delta\lambda$ in every calculation, but there is more that is to be said about it.

As precision measurements tell what is the magnetic moment correction then it is possible to see that the expected order of magnitude is obtained when the Compton wave-length correction is of the order of magnitude of the constant of fine-structure over the mass: with this foresight we may write $\delta\lambda\!=\!k\frac{\alpha}{m}$ and therefore we obtain
\begin{eqnarray}
&\vec{\mu}\!\approx\!\vec{s}\frac{q}{2m}2\left(1\!+\!\frac{1}{2}\frac{\alpha}{\pi}
\!+\!\frac{1-k}{4}\left|\frac{\alpha}{\pi}\right|^2\right)
\end{eqnarray}
in which the parameter $k$ is of the order of unity.

In the expansion $k$ begins to affect the coefficients starting from the power two: it is interesting to remark that the magnetic moment can be interpolated precisely for the electron by the parameter $k\!\approx\!2.53$ while for the muon by the parameter $k\!\approx\!-2.11$ and for the tau by the parameter $k\!\approx\!-10.33$ which are of the same magnitude.

Let us recollect now the main features: first of all, the initial relationship 
$|\vec{r}\!-\!\vec{r}\,'|\!=\!\lambda\!+\!k\frac{\alpha}{m}$ spells that the two chiral projections are separated by a distance that in average is the Compton wave-length with a correction of the order of magnitude of the leptonic classical radius, an occurrence that we find curious; in the second place, the parameter $k$ has influences that are present for the power two and higher, which means that the leptonic magnetic moment correction at the lowest-order can be explained as the electrodynamic mutual interaction between the two chiral components at the distance of the Compton wave-length plus a given correction, and that terms at higher-order would be addressed not only in terms of electrodynamic back-reaction but also in terms of this correction to the distance that separates the two chiral projections. Such an adjustable factor allows diverse fine-structure corrections for the different leptons.

We notice that the field extension diminishes when the particle mass increases and thus we speculate that more information may be available if we find some link between size and mass, maybe in terms of torsion gravity.
\section{INTERPRETATION}
To interpret what we have been doing, we may say that we have considered the lepton no longer as a point-like particle but as an extended field with an internal structure constituted by two chiral projections themselves taken as point-like particles and separated by the associated Compton wave-length; this picture may look na\"{\i}ve but it is merely the application for leptons of a picture that for hadrons is successful: as hadrons are composite of quarks and their chromodynamic interactions similarly leptons can be thought as composite of two chiral projections and their electrodynamic interactions.

Nevertheless, we have that the simpler structure of leptons compared to hadrons and the weaker coupling of photons compared to gluons are why the correction of leptons compared to hadrons is less dramatic.
\section{CONCLUSION}
In this paper, we have considered the heuristic interpretation of leptons as extended objects with an internal structure given by the two chiral projections localized in two small regions separated by the Compton wave-length of the mass of the particle, and we have taken the non-relativistic limit; we have seen that ultra-violet divergences do not necessarily arise and we have calculated the leptonic magnetic moment correction to the lowest-order, eventually speculating about possible reasons for higher-order corrections: we have remarked that leptonic magnetic moment corrections and hadronic magnetic moment anomalies may have an analogous interpretation and so they may have the very same physical meaning.

In this picture, we considered no tool whose existence might be questioned in the same way in which one may question the existence of the interaction picture through the Haag theorem: the difference compared to quantum field theory is that here leptons have an internal structure, but this comes from the fact that leptons are reducible, which is no assumption; nor is it in debate the fact that the field extension is the Compton wave-length associated to the particle mass. Hence, the correction to the gyro-magnetic factor of leptons is described in terms of the electrodynamic interaction between the two chiral projections of the field; this result has to be taken together with the fact that the hyper-fine splitting can be described in terms of a displacement in the location of the electron due to its Zitterbewegung, and with the original description of the Casimir effect as due to the retardation in van der Waals forces. These threes results describe in terms of extended fields and their retarded interactions the three effects commonly described for point particles in terms of prescriptions involving field quantization.

There is however a point that must be addressed about the hypothesis we assumed: although there is no debate around the fact that the average extension of the field be the Compton wave-length and despite that this condition is assumed systematically in quantum field theory as well, nevertheless it remains unjustified; and on the contrary we know that there are situations in which such a condition does not hold, as for the case of hadrons since their dimension is given by chromodynamic confinement.

Nevertheless, it may be that the underlying mechanism is essentially correct for leptons, quarks and hadrons, and that the differences appearing for the last instances could simply be corrections arising from strong processes.

Were this the case, then classical extended fields could replace quantization protocols; in QED, the common procedure is that of considering the particle to be point-like although quantization would give rise to a surrounding cloud of virtual bosons that makes point-like particles look like classical extended fields: but it may well be that quantum particles actually are classical extended fields.

That implementing field quantization for point-like particles may merely mean considering classical extended fields is an idea underlying in the accepted interpretation the entire framework of QED: we suggest that this is no coincidence and that it has to be taken seriously.

Retaining the description in terms of classical extended fields is theoretically simpler although it will take time for this idea to get the same QED precision.

But it may be curious to think at what might have happened if this idea came back in 1947.

\end{document}